\documentclass[usenatbib,fleqn,useAMS]{mn2e}

\usepackage{amsmath}
\usepackage{txfonts}

\renewcommand\today{30 December 2010}

\renewcommand\pi\upi
\renewcommand\partial\upartial
\newcommand\dm{\mathrm d}
\newcommand\im{\mathrm i}

\title[augmented densities of spherical systems]
{On the augmented density of a spherical anisotropic dynamic system}

\author[J.~An]
{J.~An\thanks{E-mail:~jinan@nao.cas.cn}
\\National Astronomical Observatories, Chinese Academy of Sciences,
A20 Datun Road, Chaoyang District, Beijing 100012, PR China}

\date{(draft ver.\ \today)}

\pagerange{\pageref{firstpage}--\pageref{lastpage}}\pubyear{2011}

\begin{document}
\label{firstpage}
\maketitle

\begin{abstract}
This paper presents a set of new conditions on the
augmented density of a spherical anisotropic system that is necessary
for the underlying two-integral phase-space distribution function to
be non-negative.
In particular, it is shown that the partial derivatives of the Abel
transformations of the augmented density must be non-negative.
Applied for the separable augmented densities, this recovers the result
of \citet{vHBD}.
\end{abstract}


\section{Introduction}

Recently, \citet{CM10b} raised a question whether the density
slope--anisotropy inequality $\gamma\ge2\beta$ is the necessary condition
for the consistency of the underlying two-integral phase-space
distribution function. Here, $\gamma$ is the (negative) logarithmic
density slope whereas $\beta$ is the so-called Binney anisotropy parameter.
Extending the earlier finding by \citet{AE06} that the inequality is
necessary at the centre (given a finite potential well), \citet{CM10b} have
shown that, for wide varieties of anisotropic spherical systems built
by flexible families of analytic two-integral distributions of certain
ansatz \citep[e.g.,][]{Cu91,CP95,BvH07}, the fore-mentioned inequality
holds everywhere in radial positions given that the distribution
function is also non-negative everywhere in the accessible phase space and
that the central anisotropy parameter $\beta_0$ is restricted to be
$\beta_0\le\frac12$. They also presented the equivalent condition to
the inequality for the separable augmented density, which has
subsequently proven by \citet*{vHBD} to be satisfied by such a system
if $\beta_0\le\frac12$. The latter authors have also shown that the
condition of \citet{CM10b} is not valid if $\beta_0>\frac12$,
which they have demonstrated with three counterexamples.

Here, I present new conditions on the augmented density that are
necessary for the consistency of the distribution function. It is
found that when they are applied for a separable augmented density, one
of the conditions recovers the result of \citet{vHBD}, providing an
alternative proof.

\section{augmented density for spherical systems}

According to the \citeauthor{Je15} theorem, a general spherical
dynamical system in equilibrium is described by the phase-space
distribution function (DF) of the form of $f(\mathcal E,|\bmath L|^2)$
where $\mathcal E$ and $|\bmath L|^2$ are the two classical isotropic
isolating integrals admitted by the spherical potential, namely,
$\mathcal E=\Psi-\frac12v^2$ is the specific binding energy and
$L=\sqrt{|\bmath L|^2}=rv_\mathrm t$ is the magnitude of the specific
angular momentum. Here the notation of \citet{BT} is followed so that
$\Psi$ is the relative potential above the boundary ($\Psi=-\Phi$
where $\Phi$ is the gravitational potential if the system is
infinitely extended but the potential is escapable) and the bound
particles are given by $\mathcal E>0$. In addition, $v^2=v_\mathrm t^2+
v_r^2$ where $v_\mathrm t$ and $v_r$ are the tangential and radial
velocities. This paper concerns an escapable system or a system with
a finite boundary, and therefore the local velocity
distribution always cuts off (at the escape speed) and $\mathcal E<0$
is inaccessible.\footnote{This excludes some systems such as an
isothermal sphere. Some of the following results are still valid for
such systems if they are not dependent upon the choice of the lower
boundary $\mathcal E=0$ for $\mathcal E$-integral whereas the
corresponding proper result may be addressed by choosing $\mathcal E=
-\infty$ instead.}

Given the DF, its velocity moments are given by
\begin{equation}\label{eq:pnm}\begin{split}
p_{n,m}(\Psi,r^2)
&=\iiint_{v^2\le2\Psi}\!\dm^3\!\bmath v\,v_r^{2n}v_\mathrm t^{2m}
f\bigl(\mathcal E=\Psi-\tfrac{v^2}2,L^2=r^2v_\mathrm t^2\bigr)
\\&=4\pi\!\iint_{\stackrel{v_\mathrm t\ge0,\,v_r\ge0}
{v_\mathrm t^2+v_r^2\le2\Psi}}
f\,v_r^{2n}v_\mathrm t^{2m+1}\,\dm v_\mathrm t\,\dm v_r.
\end{split}\end{equation}
Here, the zeroth moment $p_{0,0}(\Psi,r^2)$ as a bivariate function of
$\Psi$ and $r^2$ is referred to as the augmented density. Once the
potential $\Psi(r)$ is specified, the local density is found by
$\rho(r)=p_{0,0}[\Psi(r),r^2]$ whilst all the even-integral velocity
moments are given by $\langle v_r^{2n}v_\mathrm t^{2m}\rangle=
p_{n,m}/p_{0,0}$ with the specified $\Psi=\Psi(r)$. In a
self-consistent system on the other hand, the Poisson equation
with the augmented density as the source term reduces to an ordinary
differential equation on $\Psi$, and so $\rho$ becomes
entirely specified by $f$.

By changing integration variables, the augmented density
$p_{0,0}(\Psi,r^2)$ is formally expressed to be a bivariate Abel-like
integral transformation of the DF, that is,
\begin{equation}\label{eq:vmi}
p_{0,0}(\Psi,r^2)
=\frac{2\pi}{r^2}\!
\iint_{\mathcal E\ge0,\,L^2\ge0,\,\mathcal K\ge0}\!\dm\mathcal E\,\dm L^2\,
\frac{f(\mathcal E,L^2)}{\mathcal K^{1/2}}
\end{equation}
where
\begin{equation}\label{eq:kdef}
\mathcal K(\mathcal E,L^2;\Psi,r^2)
\equiv2(\Psi-\mathcal E)-\frac{L^2}{r^2}.
\end{equation}
The integral is over the triangular domain in $(\mathcal E,L^2)$
bounded by lines $\mathcal E=0$, $L^2=0$ and $\mathcal K=0$. The
last line is the same as the diagonal line given by $\mathcal E+
L^2/(2r^2)=\Psi$.
It is known that upon certain restrictions on its regularity,
equation (\ref{eq:vmi}) can be uniquely inverted at least formally
to yield the DF, $f(\mathcal E,L^2)$, provided that
$p_{0,0}(\Psi,r^2)$ is sufficiently well-behaving. Two particular
such formulae after \citet{HQ93} are provided in Appendix \ref{app:inv}.

\section{Abel transform of augmented density}

Let us consider two separate changes of integration variables in
equation (\ref{eq:vmi}), to $(Q,L^2)$ and $(\mathcal E,R^2)$,
respectively, where
$Q\equiv\mathcal E+L^2/(2r^2)$ and $R^2\equiv L^2/[2(\Psi-\mathcal E)]$.

\subsection{on the potential}

With $Q=\mathcal E+L^2/(2r^2)$, it is easy to find that the Jacobian
determinant for the coordinate change $(\mathcal E,L^2)\rightarrow
(Q,L^2)$ is simply the unity. Since we then have $\mathcal K=2(\Psi-Q)$,
the augmented density is written to be
\begin{equation}\begin{split}
p_{0,0}(\Psi,r^2)&=\frac{2\pi}{r^2}
\!\iint_{\mathcal E_{\max}(Q)\ge0,\,L^2\ge0,\,Q\le\Psi}\!\dm Q\,\dm L^2\,
\frac{f\bigl[\mathcal E_{\max}(Q),L^2\bigr]}{\sqrt{2(\Psi-Q)}}
\\&=\frac{\sqrt2\pi}{r^2}
\!\int_0^\Psi\!\frac{\dm Q}{\sqrt{\Psi-Q}}
\!\int_0^{2r^2Q}\!\dm L^2f\bigl[\mathcal E_{\max}(Q),L^2\bigr]
\end{split}\end{equation}
where $\mathcal E_{\max}(Q)\equiv Q-L^2/(2r^2)$.
Here, the inner integral in the last line is
independent of $\Psi$, and furthermore, the outer integral is in
the form of the Abel transform. Consequently, the inner integral can
be inverted from $p_{0,0}$ by means of the inverse Abel transform.
That is to say,
\begin{equation}\boxed{
\frac\partial{\partial Q}\!
\int_0^Q\!\frac{p_{0,0}(\Psi,r^2)\,\dm\Psi}{\sqrt{Q-\Psi}}
=\frac{\sqrt2\pi^2}{r^2}\!
\int_0^{2r^2Q}\!\dm L^2f\bigl[\mathcal E_{\max}(Q),L^2\bigr]}.
\end{equation}

Next, we can show that
\begin{multline}
\frac\partial{\partial r^2}\!
\int_0^{2r^2Q}\!\dm L^2f\bigl[\mathcal E_{\max}(Q),L^2\bigr]
\\=2Qf(0,2r^2Q)+
\int_0^{2r^2Q}\!\dm L^2f^{(1,0)}\bigl[\mathcal E_{\max}(Q),L^2\bigr]\,
\frac{L^2}{2r^4}
\end{multline}
where $f^{(1,0)}(\mathcal E,L^2)\equiv\partial f/\partial\mathcal E$.
However, we also find that
\begin{equation}\begin{split}
\frac\partial{\partial Q}\!
&\int_0^{2r^2Q}\!\dm L^2L^2f\bigl[\mathcal E_{\max}(Q),L^2\bigr]
\\&=4r^4Qf(0,2r^2Q)+
\int_0^{2r^2Q}\!\dm L^2L^2f^{(1,0)}\bigl[\mathcal E_{\max}(Q),L^2\bigr]
\\&=2r^4\frac\partial{\partial r^2}\!
\int_0^{2r^2Q}\!\dm L^2f\bigl[\mathcal E_{\max}(Q),L^2\bigr].
\end{split}\end{equation}
Hence,
\begin{equation}\boxed{
\frac\partial{\partial r^2}\!\left\lgroup
r^2\!\int_0^Q\!\frac{p_{0,0}(\Psi,r^2)\,\dm\Psi}{\sqrt{Q-\Psi}}\right\rgroup
=\frac{\pi^2}{\sqrt2r^4}\!
\int_0^{2r^2Q}\!\dm L^2L^2f\bigl[\mathcal E_{\max}(Q),L^2\bigr]}.
\end{equation}
Strictly, the argument so far only indicates that the difference
between the left- and right-hand sides is independent of $Q$, but both
vanish if $Q=0$ and thus the equality holds.

\subsection{on the radial distance}

Given $R^2=L^2/[2(\Psi-\mathcal E)]$, we first find the Jacobian determinant
\begin{equation}
\left|\frac{\partial(\mathcal E,L^2)}{\partial(\mathcal E,R^2)}\right|
=2(\Psi-\mathcal E)
\end{equation}
for the coordinate change $(\mathcal E,L^2)\rightarrow(\mathcal E, R^2)$
whilst we have that
\begin{equation}
\mathcal K=2(\Psi-\mathcal E)-\frac{2R^2(\Psi-\mathcal E)}{r^2}
=2(\Psi-\mathcal E)\,\frac{r^2-R^2}{r^2}.
\end{equation}
Then, equation (\ref{eq:vmi}) reduces to
\begin{equation}
p_{0,0}(\Psi,r^2)
=\frac{2^{3/2}\pi}r\!
\int_0^{r^2}\!\dm R^2\!\int_0^\Psi\!\dm\mathcal E
\frac{\sqrt{\Psi-\mathcal E}}{\sqrt{r^2-R^2}}
f\bigl[\mathcal E,L^2_{\max}(R^2)\bigr].
\end{equation}
Here $L^2_{\max}(R^2)\equiv2R^2(\Psi-\mathcal E)$,
which does not depend on $r^2$. The
integral is now over the rectangular region and so the double integral
can be performed in any order given the absolutely integrable integrand.
In addition, the $R^2$-integral is again an Abel transform and its
inverse transform results in
\begin{equation}\boxed{
\frac\partial{\partial R^2}\!
\int_0^{R^2}\!\frac{rp_{0,0}(\Psi,r^2)\,\dm r^2}{\sqrt{R^2-r^2}}
=2^{3/2}\pi^2\!\int_0^\Psi\!\dm\mathcal E\sqrt{\Psi-\mathcal E}\,
f\bigl[\mathcal E,L^2_{\max}(R^2)\bigr]}.
\end{equation}

Next we note that
\begin{equation}
\frac r{\sqrt{R^2-r^2}}
=\frac{R^2}{r\sqrt{R^2-r^2}}
-\frac{\sqrt{R^2-r^2}}r,
\end{equation}
and therefore (provided that $p_{0,0}\dm r$ is integrable over $r=0$)
\begin{equation}\label{eq14}\begin{split}
\frac\partial{\partial R^2}\!
&\int_0^{R^2}\!\frac{rp_{0,0}(\Psi,r^2)\,\dm r^2}{\sqrt{R^2-r^2}}
\\&=\frac\partial{\partial R^2}\biggl[R^2\!
\int_0^{R^2}\!\frac{p_{0,0}(\Psi,r^2)\,\dm r^2}{r\sqrt{R^2-r^2}}\biggr]
-\tfrac12\!
\int_0^{R^2}\!\frac{p_{0,0}(\Psi,r^2)\,\dm r^2}{r\sqrt{R^2-r^2}}
\\&=\Bigl(\tfrac12+R^2\frac\partial{\partial R^2}\Bigr)
\!\int_0^{R^2}\!
\frac{p_{0,0}(\Psi,r^2)\,\dm r^2}{r\sqrt{R^2-r^2}}.
\end{split}\end{equation}
On the other hand,
\begin{multline}\label{eq15}
\frac\partial{\partial\Psi}\!
\int_0^\Psi\!\dm\mathcal E\sqrt{\Psi-\mathcal E}\,
f\bigl[\mathcal E,L^2_{\max}(R^2)\bigr]
=\tfrac12\!\int_0^\Psi\!\dm\mathcal E
\frac{f\bigl[\mathcal E,L^2_{\max}(R^2)\bigr]}
{\sqrt{\Psi-\mathcal E}}
\\+2R^2\!\int_0^\Psi\!\dm\mathcal E\sqrt{\Psi-\mathcal E}\,
f^{(0,1)}\bigl[\mathcal E,L^2_{\max}(R^2)\bigr]
\end{multline}
where $f^{(0,1)}(\mathcal E,L^2)\equiv\partial f/\partial L^2$
assuming $\lim_{L\rightarrow0}Lf(\mathcal E,L^2)=0$.
However, since
\begin{equation}
\frac\partial{\partial R^2}\!\left\lgroup
\frac{f\bigl[\mathcal E,L^2_{\max}(R^2)\bigr]}
{\sqrt{\Psi-\mathcal E}}\right\rgroup
=2\sqrt{\Psi-\mathcal E}\,
f^{(0,1)}\bigl[\mathcal E,L^2_{\max}(R^2)\bigr],
\end{equation}
equation (\ref{eq15}) further reduces to
\begin{multline}\label{eq17}
\frac\partial{\partial\Psi}\!
\int_0^\Psi\!\dm\mathcal E\sqrt{\Psi-\mathcal E}\,
f\bigl[\mathcal E,L^2_{\max}(R^2)\bigr]
\\=\Bigl(\tfrac12+R^2\frac\partial{\partial R^2}\Bigr)
\!\int_0^\Psi\!\dm\mathcal E
\frac{f\bigl[\mathcal E,L^2_{\max}(R^2)\bigr]}
{\sqrt{\Psi-\mathcal E}},
\end{multline}
which is in fact valid provided that $f\dm L^2/L$ is integrable
over $L^2=0$ (c.f., Appendix~\ref{appC}).
Consequently, we finally find that
\begin{equation}\label{eq18}\boxed{
\frac\partial{\partial\Psi}\!\int_0^{R^2}\!
\frac{p_{0,0}(\Psi,r^2)\,\dm r^2}{r\sqrt{R^2-r^2}}
=2^{3/2}\pi^2\!\int_0^\Psi\!\dm\mathcal E
\frac{f\bigl[\mathcal E,L^2_{\max}(R^2)\bigr]}
{\sqrt{\Psi-\mathcal E}}}.
\end{equation}

\section{Consistency of Distribution Function}

The results so far are summarized as follows. If we define the Abel
transformations of the augmented density as
\begin{equation}\label{eq:abts}\begin{split}
\hat\rho(\Psi,r^2)&\equiv
\frac1\pi\!
\int_0^\Psi\!\frac{p_{0,0}(Q,r^2)\,\dm Q}{\sqrt{\Psi-Q}};
\\
\bar\rho(\Psi,r^2)&\equiv
\frac1\pi\!
\int_0^{r^2}\!\frac{p_{0,0}(\Psi,R^2)\,\dm R^2}{R\sqrt{r^2-R^2}};
\\
\tilde\rho(\Psi,r^2)&\equiv
\frac1\pi\!
\int_0^{r^2}\!\frac{Rp_{0,0}(\Psi,R^2)\,\dm R^2}{\sqrt{r^2-R^2}},
\end{split}\end{equation}
we find that applying the transform on equation (\ref{eq:vmi}) results in
\begin{equation}\label{eq:rhot}\begin{split}
\hat\rho(\Psi,r^2)&=
\frac{\sqrt2\pi}{r^2}\!
\iint_{\mathcal E\ge0,\,L^2\ge0,\,\mathcal K\ge0}\!
\dm\mathcal E\,\dm L^2f(\mathcal E,L^2);
\\
\bar\rho(\Psi,r^2)&=
\frac{2\pi}r\!
\iint_{\mathcal E\ge0,\,L^2\ge0,\,\mathcal K\ge0}\!
\dm\mathcal E\,\dm L^2\frac{f(\mathcal E,L^2)}L;
\\
\tilde\rho(\Psi,r^2)&=
\sqrt2\pi\!
\iint_{\mathcal E\ge0,\,L^2\ge0,\,\mathcal K\ge0}\!
\dm\mathcal E\,\dm L^2\frac{f(\mathcal E,L^2)}{\sqrt{\Psi-\mathcal E}}.
\end{split}\end{equation}
The calculations are performed by interchanging orders of
integrations (so that $Q$- and $R^2$-integrations of the transforms
become the inner-most ones) and carrying out the $Q$- and
$R^2$-integrations of the transforms explicitly. The details are given
in Appendix \ref{app} \citep[see also][]{Qi93}.
Note that the middle equation
(\ref{eq:rhot}) is also identifiable to eq.~(C10) of \citet{HQ93}.

The partial derivatives of equations (\ref{eq:rhot}) are related to
simple linear integrals of the DF along the diagonal line given by
$\mathcal E+L^2/(2r^2)=\Psi$. In particular,
\begin{equation}\label{eq:main}\begin{split}
&\frac{\partial\hat\rho(\Psi,r^2)}{\partial\Psi}
=\frac{\sqrt2\pi}{r^2}\!\int_0^{2r^2\Psi}\!\dm L^2
f\Bigl(\Psi-\frac{L^2}{2r^2},L^2\Bigr);
\\&
\frac{\partial[r^2\hat\rho(\Psi,r^2)]}{\partial r^2}
=\frac\pi{\sqrt2r^4}\!\int_0^{2r^2\Psi}\!\dm L^2L^2
f\Bigl(\Psi-\frac{L^2}{2r^2},L^2\Bigr);
\\&
\frac{\partial\bar\rho(\Psi,r^2)}{\partial\Psi}
=\frac{2\pi}r\!\int_0^{2r^2\Psi}\!\frac{\dm L^2}L
f\Bigl(\Psi-\frac{L^2}{2r^2},L^2\Bigr);
\\&
\frac{\partial\tilde\rho(\Psi,r^2)}{\partial r^2}
=\frac\pi{r^3}\!\int_0^{2r^2\Psi}\!\dm L^2L\,
f\Bigl(\Psi-\frac{L^2}{2r^2},L^2\Bigr).
\end{split}\end{equation}
%
%
%
Here each integral may alternatively expressed through
\begin{multline}\label{eq:altint}
\int_0^{2r^2\Psi}\!\dm L^2L^{2m}
f\Bigl(\Psi-\frac{L^2}{2r^2},L^2\Bigr)
\\=(2r^2)^{m+1}\!\int_0^\Psi\!\dm\mathcal E\,(\Psi-\mathcal E)^m
f\bigl[\mathcal E,2r^2(\Psi-\mathcal E)\bigr].
\end{multline}
%
This is
verified by changing the dummy integration variables. The last
line of equations (\ref{eq:main}) may be expressed in terms of
$\bar\rho$ because
\begin{equation}
\frac{\partial\tilde\rho(\Psi,r^2)}{\partial r^2}
=\Bigl(\tfrac12+r^2\frac\partial{\partial r^2}\Bigr)\ \bar\rho(\Psi,r^2)
=r\frac{\partial[r\bar\rho(\Psi,r^2)]}{\partial r^2},
\end{equation}
which may be shown with equation (\ref{eq14}).

Since the minimal consistency of the DF indicates that $f\ge0$ in the
accessible region of the phase space and also $L^2\ge0$ and
$\mathcal E\le\Psi$, equations (\ref{eq:main}) are all non-negative,
provided that they are indeed well-defined. In fact, $\hat\rho$ and
$\tilde\rho$ are well-defined given that the DF is integrable --
which further implies that $p_{0,0}(\Psi,r^2)\dm\Psi$ and
$p_{0,0}(\Psi,r^2)\dm r^2$ are integrable over $\Psi=0$ and $r^2=0$,
respectively. However, the definition of $\bar\rho$ and subsequently
the third line of equations (\ref{eq:main}) require
$f(\mathcal E,L^2)\dm L^2/L$
to be integrable over the region containing $L^2=0$, which is
a strictly stronger condition than the mere integrability of the DF.
For instance,
if $f(\mathcal E,L^2)=L^{-2\beta}g(\mathcal E)$ and
$p_{0,0}(\Psi,r^2)=r^{-2\beta}A(\Psi)$ with $\frac12\le\beta(<1)$,
it is clear that $\bar\rho$ cannot be defined as in equation
(\ref{eq:abts}) whilst the right-hand side of the third line of
equations (\ref{eq:main}) diverges.

\subsection{separable augmented density}

Next, we consider the particular cases for which the $\Psi$- and
$r^2$-dependences of the augmented density are multiplicatively
separable, that is, $p_{0,0}(\Psi,r^2)=A(\Psi)B(r^2)$. It is
a fairly straightforward exercise to establish \citep[c.f.,][]{De86}
\begin{gather}\label{eq:den}
\frac{\partial p_{1,0}}{\partial\Psi}=p_{0,0}
\quad\Rightarrow\
p_{1,0}(\Psi,r^2)=\int_0^\Psi\!p_{0,0}(Q,r^2)\,\dm Q,
\intertext{and}
\frac{\partial(r^2p_{1,0})}{\partial r^2}
=\tfrac12\,p_{0,1}
\quad\Rightarrow\
1-\frac{p_{0,1}}{2p_{1,0}}
=-\frac{\partial\log p_{1,0}}{\partial\log r^2},
\end{gather}
from equation (\ref{eq:pnm}). The last result is also directly related
to the Binney anisotropy parameter \citep{BM82}, i.e.,
\begin{equation}\label{eq:beta}
\beta(r)=1-\frac{\langle v_\mathrm t^2\rangle}{2\langle v_r^2\rangle}
=1-\frac{p_{0,1}[\Psi(r),r^2]}{2p_{1,0}[\Psi(r),r^2]}
=\left.-\frac{\partial\log p_{1,0}}{\partial\log r^2}\right|_{\Psi(r),r^2}.
\end{equation}
Therefore, with a separable augmented density, we further have
\begin{displaymath}
p_{1,0}(\Psi,r^2)=B(r^2)\int_0^\Psi\!A(Q)\,\dm Q,
\end{displaymath}
which is also separable. In addition,
\begin{displaymath}
\beta=-\frac{\dm\log B(r^2)}{\dm\log r^2}
\,;\qquad
\log B=-\int\frac{2\beta(r)}r\dm r,
\end{displaymath}
which is thus completely specified (up to a constant) given the
anisotropy parameter.

The first and last of equations (\ref{eq:main}) being non-negative for
a separable augmented density then indicates that
\begin{equation}\label{eq:sep1}
\frac{\dm\hat A(\Psi)}{\dm\Psi}\ge0
\,;\qquad
\frac{\dm\tilde B(r^2)}{\dm r^2}\ge0,
\end{equation}
where
\begin{equation}
\hat A(\Psi)\equiv\int_0^\Psi\!\frac{A(Q)\,\dm Q}{\sqrt{\Psi-Q}}
\,;\qquad
\tilde B(r^2)\equiv\int_0^{r^2}\!\frac{RB(R^2)\,\dm R^2}{\sqrt{r^2-R^2}}.
\end{equation}
Here we have also used $A(\Psi)\ge0$ and $B(r^2)\ge0$, which are
direct consequences of the DF being non-negative.
It is also obvious
that $\hat A(\Psi)\ge0$, and so we find from the second line
of equation (\ref{eq:main}) that
\begin{equation}\label{eq29}
\frac{\dm[r^2B(r^2)]}{\dm r^2}=(1-\beta)B\ge0.
\end{equation}
Given that $B(r^2)\ge0$, this is equivalent to $\beta\le1$, which is
physically obvious. None the less, it is a nontrivial necessary
condition on $B(r^2)$ for the consistency of the corresponding DF.
Furthermore, we also find that
\begin{equation}
\frac\dm{\dm r^2}\!\int_0^{r^2}\!\frac{RB(R^2)\,\dm R^2}{\sqrt{r^2-R^2}}
=\frac1{r^2}\!\int_0^{r^2}\!\frac{\dm[R^2B(R^2)]}{\dm R^2}
\frac{R\,\dm R^2}{\sqrt{r^2-R^2}},
\end{equation}
and thus equation (\ref{eq29}) is in fact a sufficient condition
for the second part of equation (\ref{eq:sep1}).
Finally, let us suppose that
\begin{equation}\label{eq:barb}
\bar B(r^2)\equiv\int_0^{r^2}\!\frac{B(R^2)\,\dm R^2}{R\sqrt{r^2-R^2}}
=\int_0^\pi\!B(r^2\!\sin^2\!\tfrac\phi2)\,\dm\phi
\end{equation}
converges (and so is well-defined). Then since it is also
non-negative, the non-negativity of the DF indicates that
\begin{equation}\label{eq:sep2}
\frac{\dm A(\Psi)}{\dm\Psi}\ge0
\,;\qquad
\frac{\dm[r\bar B(r^2)]}{\dm r^2}
=\frac1r\frac{\dm\tilde B(r^2)}{\dm r^2}\ge0.
\end{equation}
Here we note that
\begin{equation}
\frac{\dm\hat A(\Psi)}{\dm\Psi}
=\frac{A(0)}{\sqrt\Psi}+\int_0^\Psi\!\frac{\dm Q}{\sqrt{\Psi-Q}}
\frac{\dm A(Q)}{\dm Q},
\end{equation}
and so equation (\ref{eq:sep2}) is also a sufficient condition for
equation (\ref{eq:sep1}). However, equation (\ref{eq:sep2}) is the
necessary condition for the consistency of the DF provided that
$\bar B(r^2)$ is well-defined in equation (\ref{eq:barb}) whereas
equation (\ref{eq:sep1}) is the necessary condition for the
non-negativity of any integrable DF that produces a separable
augmented density.

The partial result of \citet{vHBD} is also deduced from these. If
$\beta_0=\lim_{r\rightarrow0}\beta(r)$, this implies that $B(r^2)\sim
r^{-2\beta_0}$ as $r\rightarrow0$. Hence, if $2\beta_0<1$, then
$\bar B$ converges and is well-defined. Consequently, equation
(\ref{eq:sep2}), in particular, $\dm A/\dm\Psi\ge0$ is the necessary
condition for the consistency of the DF. Following \citet{CM10b}, this
is also equivalent to the density slope--anisotropy inequality
($\gamma\ge2\beta$) for the system with a separable augmented density.
The result here is however not directly applicable for the case that
$\beta_0=\frac12$, for which equation (\ref{eq:barb}) diverges.

\section*{acknowledgments}

Large parts of this work were done during the author's two-week
visit to the IoA (Cambridge) during October 2010, which was in part
supported by the National Natural Science Foundation of China (NSFC)
Research Fund for International Young Scientist as well as
the IoA's visitor grant.
The author thanks Wyn Evans in particular for lending
his copy of Dr Edward Enping Qian's PhD Thesis, which was invaluable
for completion of this work, and also acknowledges Martin Smith
for pointing to \citet{vHBD}, which was the immediate impetus of the
current paper. The author also thanks Luca Ciotti and Maarten Baes
for the comments made on the original version of this paper,
and the anonymous referee for helping to clarify the assumptions
behind the result. The author is supported
by the Chinese Academy of Sciences (CAS) Fellowships for Young
International Scientists, Grant~No.:~2009Y2AJ7.

\begin{appendix}

\section{Abel transformation of 
$\bmath{\lowercase{p_{n,m}}(\Psi,\lowercase{r^2})}$}
\label{app}

It is easy to establish that the moment function $p_{n,m}(\Psi,r^2)$
defined in equation (\ref{eq:pnm}) is all expressed as an integral
transformation of the DF in $(\mathcal E,L^2)$-space, i.e.,
\begin{equation}\label{eq:pnmi}
r^{2m}p_{n,m}(\Psi,r^2)
=\frac{2\pi}{r^2}
\!\iint_\mathcal T\!
\dm\mathcal E\,\dm L^2\mathcal K^{n-\frac12}L^{2m}f(\mathcal E,L^2)
\end{equation}
where the integral is over the domain in the $(\mathcal E,L^2)$
defined to be
\begin{displaymath}
\mathcal T=\bigl\{\,(\mathcal E,L^2)\,|\
\mathcal E\ge0,L^2\ge0,\mathcal K\ge0\,\bigr\}.
\end{displaymath}
Here, $L^{2m}=(L^2)^m$ is itself a function of only $L^2$ and so we
may set $m=0$ without any loss of generality in the following
discussion -- formally, $F(\mathcal E,L^2)=L^{2m}f(\mathcal E,L^2)$, etc.

Next, we define the Abel transformations of equation (\ref{eq:pnmi}),
\begin{equation}\begin{split}
\hat p_n(\Psi,r^2)&\equiv
\frac1\pi\!
\int_0^\Psi\!\frac{p_{n,0}(Q,r^2)\,\dm Q}{(\Psi-Q)^{1/2}};
\\
\bar p_n(\Psi,r^2)&\equiv
\frac1\pi\!
\int_0^{r^2}\!\frac{p_{n,0}(\Psi,R^2)\,\dm R^2}{R(r^2-R^2)^{1/2}};
\\
\tilde p_n(\Psi,r^2)&\equiv
\frac1\pi\!
\int_0^{r^2}\!\frac{R^{2n+1}p_{n,0}(\Psi,R^2)\,\dm R^2}{(r^2-R^2)^{1/2}}.
\end{split}\end{equation}
Interchanging the orders of integrations indicates that
\begin{equation}\begin{split}
\hat p_n(\Psi,r^2)
&=\frac2{r^2}\!\iint_\mathcal T\!\dm\mathcal E\,\dm L^2f\!
\int_{\mathcal E+\frac{L^2}{2r^2}}^\Psi\!
\frac{[\mathcal K(Q)]^{n-\frac12}\dm Q}{(\Psi-Q)^{1/2}};
\\\bar p_n(\Psi,r^2)&=2\!\iint_\mathcal T\!
\dm\mathcal E\,\dm L^2f\!
\int_{\frac{L^2}{2(\Psi-\mathcal E)}}^{r^2}\!
\frac{[\mathcal K(R^2)]^{n-\frac12}\dm R^2}{R^3(r^2-R^2)^{1/2}};
\\\tilde p_n(\Psi,r^2)&=2\!\iint_\mathcal T\!
\dm\mathcal E\,\dm L^2f\!
\int_{\frac{L^2}{2(\Psi-\mathcal E)}}^{r^2}\!
\frac{R^{2n-1}[\mathcal K(R^2)]^{n-\frac12}\dm R^2}{(r^2-R^2)^{1/2}},
\end{split}\end{equation}
where $\mathcal K(Q)\equiv\mathcal K(\mathcal E,L^2;Q,r^2)$ and
$\mathcal K(R^2)\equiv\mathcal K(\mathcal E,L^2;\Psi,R^2)$ with
$\mathcal K$ being defined in equation (\ref{eq:kdef}). The inner-most
integrals can be evaluated analytically. In particular,
\begin{equation}\begin{split}
\int_{\mathcal E+\frac{L^2}{2r^2}}^\Psi\!
&\frac{[\mathcal K(Q)]^{n-\frac12}\dm Q}{(\Psi-Q)^{1/2}}
=\frac{\mathcal K^n}{\sqrt 2}\,\mathrm B\bigl(n+\tfrac12,\tfrac12\bigr);
\\\int_{\frac{L^2}{2(\Psi-\mathcal E)}}^{r^2}\!
&\frac{[\mathcal K(R^2)]^{n-\frac12}\dm R^2}{R^3(r^2-R^2)^{1/2}}
=\frac{\mathcal K^n}{rL}\,\mathrm B\bigl(n+\tfrac12,\tfrac12\bigr);
\\\int_{\frac{L^2}{2(\Psi-\mathcal E)}}^{r^2}\!
&\frac{R^{2n-1}[\mathcal K(R^2)]^{n-\frac12}\dm R^2}{(r^2-R^2)^{1/2}}
=\frac{r^{2n}\mathcal K^n}{\sqrt{2(\Psi-\mathcal E)}}\,
\mathrm B\bigl(n+\tfrac12,\tfrac12\bigr),
\end{split}\end{equation}
where $\mathcal K\equiv\mathcal K(\mathcal E,L^2;\Psi,r^2)$, and
\begin{displaymath}
\mathrm B\bigl(n+\tfrac12,\tfrac12\bigr)\equiv
\int_0^1\!\frac{x^{n-\frac12}\dm x}{(1-x)^{1/2}}
=\frac{\Gamma(n+\frac12)\Gamma(1/2)}{\Gamma(n+1)}
=\frac{(1/2)_n\pi}{n!}.
\end{displaymath}
The last equality is valid for a non-negative integer $n$, 
and $(a)_n\equiv\prod_{i=0}^{n-1}(a+i)$ is the Pochhammer symbol.
This may be proven by the changes of variables given by
\begin{equation}
\hat x(Q)=\frac{\mathcal K(Q)}{\mathcal K};\quad
\bar x(R^2)=\frac{\mathcal K(R^2)}{\mathcal K};\quad
\tilde x(R^2)=\frac{R^2\mathcal K(R^2)}{r^2\mathcal K},
\end{equation}
respectively. Finally, gathering the results so far results in
\begin{equation}\begin{split}
\hat p_n(\Psi,r^2)&
=\frac{(1/2)_n}{n!}\frac{\sqrt 2\pi}{r^2}\!
\iint_\mathcal T\!\dm\mathcal E\,\dm L^2
\mathcal K^nf(\mathcal E,L^2);\\
\bar p_n(\Psi,r^2)&
=\frac{(1/2)_n}{n!}\frac{2\pi}r\!
\iint_\mathcal T\!\dm\mathcal E\,\dm L^2
\mathcal K^n\frac{f(\mathcal E,L^2)}L;\\
\tilde p_n(\Psi,r^2)&
=\frac{(1/2)_n}{n!}\sqrt 2\pi r^{2n}\!
\iint_\mathcal T\!\dm\mathcal E\,\dm L^2
\mathcal K^n\frac{f(\mathcal E,L^2)}{(\Psi-\mathcal E)^{1/2}}.
\end{split}\end{equation}

\section{Hunter-Qian inversion for
$\bmath{\lowercase{f}(\mathcal E,L^2)}$}
\label{app:inv}

The Abel transformations of the augmented density examined in
this paper indicate that the particular problem of inverting
the augmented density for the DF, $f(\mathcal E,L^2)$ is also
closely related to finding the (even part of) two-integral
distribution function $f^+(\mathcal E,J_z^2)$ from the axisymmetric
density $\rho[\Psi(R^2,z^2),R^2]$, as the integral transform that
relates them can be formally identified with equation (\ref{eq:rhot}).
\citet{HQ93} had presented
the complete solution for this latter problem by means of the
complex contour integral, and also discussed its relation to
the spherical anisotropic case. Although they did not provide
explicitly the corresponding general inversion formulae
for anisotropic spherical systems, they are in fact no more
complicated than the axisymmetric case \citep[eq.~3.1]{HQ93}.
That is to say \citep[see also][]{Qi93},
\begin{equation}\label{eq:hqci}\begin{split}
f(\mathcal E,L^2)
&=\frac1{2^{5/2}\pi^2\im}\,
\frac\partial{\partial\mathcal E}\!
\int_{\mathcal E_{\min}}^{(\Psi_\mathrm{env})+}\!
\frac{\dm\mathcal Z}{\mathcal Z-\mathcal E}\,
\hat{\mathcal P}
\biggl[\mathcal Z,\frac{L^2}{2(\mathcal Z-\mathcal E)}\biggr]
\\&=\frac1{2^{5/2}\pi^2\im}\,
\frac\partial{\partial\mathcal E}\!
\int_{\mathcal E_{\min}}^{(\Psi_\mathrm{env})+}\!
\frac{\dm\mathcal Z}{(\mathcal Z-\mathcal E)^{1/2}}\,
\bar{\mathcal P}
\biggl[\mathcal Z,\frac{L^2}{2(\mathcal Z-\mathcal E)}\biggr]
\end{split}\end{equation}
where
\begin{displaymath}\begin{split}
\hat{\mathcal P}(\Psi,r^2)
&=\frac{\partial\hat\rho(\Psi,r^2)}{\partial\Psi}
=\frac1\pi\frac\partial{\partial\Psi}\!
\int_0^\Psi\!\frac{p_{0,0}(Q,r^2)\,\dm Q}{(\Psi-Q)^{1/2}}
\\
\bar{\mathcal P}(\Psi,r^2)
&=\frac{\partial\bar\rho(\Psi,r^2)}{\partial\Psi}
=\frac1\pi\frac\partial{\partial\Psi}\!
\int_0^{r^2}\!\frac{p_{0,0}(\Psi,R^2)\,\dm R^2}{R(r^2-R^2)^{1/2}},
\end{split}\end{displaymath}
and $\hat\rho(\Psi,r^2)$ and $\bar\rho(\Psi,r^2)$ are as defined
in equation (\ref{eq:abts}).
Here, the outer integrals of equation (\ref{eq:hqci}) are contour
integrals in the complex $\mathcal Z$-plane whose path is given
such that
it starts from $\mathcal Z=\mathcal E_{\min}$
to the below of the real axis, winds along
the positive (counterclockwise) orientation, crosses the real axis upwards
(from the negative to positive imaginary) at the right side of
$\mathcal Z=\Psi_\mathrm{env}(\mathcal E)$, and returns to
$\mathcal Z=\mathcal E_{\min}$ from
the positive imaginary side. (Note that, given that
$\mathcal E<0$ is inaccessible, $\mathcal E_{\min}=0$.)
For more details including the definition
of $\Psi_\mathrm{env}(\mathcal E)$, please refer \citet{HQ93} or
\citet{Qi93}. We note that the formula
reduces to the \citeauthor{Cu91} formula if the $r^2$-dependence of
the augmented density is given by a pure power law [i.e.,
$p_{0,0}(\Psi,r^2)=r^{-2\beta}A(\Psi)$]. Of course, this further
reduces to the \citeauthor{Ed16} formula if $\beta=0$.

\vfill
\section{alternative derivation of e\lowercase{q.\ref{eq17}}}
\label{appC}

In equation (\ref{eq15}), we have assumed
$\lim_{L\rightarrow0}Lf(\mathcal E,L^2)=0$ in order to omit the term
$\lim_{\Psi\rightarrow\mathcal E}\sqrt{\Psi-\mathcal E}
f[\mathcal E,L_{\max}^2(R^2)]$ in its right-hand side. However, this
assumption is only incidental here and the result in equation
(\ref{eq17}) and subsequently equation (\ref{eq18}) are valid provided
that the integrals there are well-defined and differentiable. This may
be argued based on the fact that equation (\ref{eq18}) is rederived
as the third line of equations (\ref{eq:main}) by differentiating
equation (\ref{eq:rhot}) directly and using equation (\ref{eq:altint}).
Alternatively, using equation (\ref{eq:altint}), we first find that
\begin{multline*}
\int_0^\Psi\!\dm\mathcal E\sqrt{\Psi-\mathcal E}\,
f\bigl[\mathcal E,2R^2(\Psi-\mathcal E)\bigr]
\\=\frac1{(2R^2)^{3/2}}\!\int_0^{2R^2\Psi}\!\dm L^2L
f\Bigl(\Psi-\frac{L^2}{2R^2},L^2\Bigr);
\end{multline*}
\begin{displaymath}
\int_0^\Psi\!\dm\mathcal E
\frac{f\bigl[\mathcal E,2R^2(\Psi-\mathcal E)\bigr]}
{\sqrt{\Psi-\mathcal E}}
=\frac1{(2R^2)^{1/2}}\!\int_0^{2R^2\Psi}\!\frac{\dm L^2}L
f\Bigl(\Psi-\frac{L^2}{2R^2},L^2\Bigr).
\end{displaymath}
Then, the direct differentiations indicate that
\begin{multline*}
\frac\partial{\partial\Psi}\!
\int_0^\Psi\!\dm\mathcal E\sqrt{\Psi-\mathcal E}\,
f\bigl[\mathcal E,2R^2(\Psi-\mathcal E)\bigr]
\\=\Psi^{1/2}f(0,2R^2\Psi)
+\frac1{(2R^2)^{3/2}}\!\int_0^{2R^2\Psi}\!\dm L^2L
f^{(1,0)}\Bigl(\Psi-\frac{L^2}{2R^2},L^2\Bigr);
\end{multline*}
and
\begin{multline*}
\frac\partial{\partial R^2}\!\left\lgroup
R\!\int_0^\Psi\!\dm\mathcal E
\frac{f\bigl[\mathcal E,2R^2(\Psi-\mathcal E)\bigr]}
{\sqrt{\Psi-\mathcal E}}\right\rgroup
\\=\frac{\Psi^{1/2}}Rf(0,2R^2\Psi)
+\frac1{2^{3/2}R^4}\!\int_0^{2R^2\Psi}\!\dm L^2
\frac{L^2}L
f^{(1,0)}\Bigl(\Psi-\frac{L^2}{2R^2},L^2\Bigr).
\end{multline*}
That is to say,
\begin{multline*}
\frac\partial{\partial R^2}\!\left\lgroup
R\!\int_0^\Psi\!\dm\mathcal E
\frac{f\bigl[\mathcal E,2R^2(\Psi-\mathcal E)\bigr]}
{\sqrt{\Psi-\mathcal E}}\right\rgroup
\\=\frac1R\frac\partial{\partial\Psi}\!
\int_0^\Psi\!\dm\mathcal E\sqrt{\Psi-\mathcal E}\,
f\bigl[\mathcal E,2R^2(\Psi-\mathcal E)\bigr],
\end{multline*}
which is equivalent to equation (\ref{eq17}).

\end{appendix}

\label{lastpage}

\end{document}